\documentclass[pdflatex,sn-mathphys,Numbered]{sn-jnl}

\usepackage{graphicx}%
\usepackage{multirow}%
\usepackage{amsmath,amssymb,amsfonts}%
\usepackage{amsthm}%
\usepackage{mathrsfs}%
\usepackage[title]{appendix}%
\usepackage{xcolor}%
\usepackage{textcomp}%
\usepackage{manyfoot}%
\usepackage{booktabs}%
\usepackage{listings}%

\usepackage[capitalise]{cleveref}
\usepackage[caption=false]{subfig}

\usepackage{xfrac}
\usepackage{xcolor}
\usepackage{placeins}

\theoremstyle{thmstyleone}%
\theoremstyle{thmstyletwo}%

\theoremstyle{thmstylethree}%

\unnumbered

\begin{document}

\title[Defect trapping in Ni--Mo alloy corrosion]{Role of defect trapping in the enhanced corrosion resistance of Ni--Mo alloys with yttrium and yttrium oxide additions}


\author*[1]{\fnm{Jia-Hong} \sur{Ke}}\email{jiahong.ke@inl.gov}
\author[2]{\fnm{Xinchang} \sur{Zhang}}
\author[3]{\fnm{Jingfan} \sur{Yang}}
\author[4]{\fnm{Qiufeng} \sur{Yang}}
\author[4]{\fnm{Ruchi} \sur{Gakhar}}
\author*[3]{\fnm{Rongjie} \sur{Song}}\email{rongjie.song@inl.gov}

\affil*[1]{\orgdiv{Computational Mechanics and Materials Department}, \orgname{Idaho National Laboratory}, \orgaddress{\city{Idaho Falls}, \postcode{83415}, \state{ID}, \country{USA}}}

\affil*[2]{\orgdiv{Materials Science and Manufacturing Department}, \orgname{Idaho National Laboratory}, \orgaddress{\city{Idaho Falls}, \postcode{83415}, \state{ID}, \country{USA}}}

\affil*[3]{\orgdiv{Irradiated Fuels and Materials Department}, \orgname{Idaho National Laboratory}, \orgaddress{\city{Idaho Falls}, \postcode{83415}, \state{ID}, \country{USA}}}

\affil*[4]{\orgdiv{Advanced Technology of Molten Salts Department}, \orgname{Idaho National Laboratory}, \orgaddress{\city{Idaho Falls}, \postcode{83415}, \state{ID}, \country{USA}}}


\abstract{Corrosion of structural alloys remains a critical challenge for advanced molten salt reactors, where aggressive fluoride salts promote continuous dissolution at the salt--metal interface. While most mitigation strategies emphasize thermodynamic stability, kinetic control of atomic transport has received comparatively less attention. Here, we propose a corrosion mitigation strategy based on diffusion control and vacancy trapping by dilute yttrium (Y) solutes and yttrium oxide (Y$_2$O$_3$) dispersions in Ni alloys. A combined computational and experimental approach integrates density functional theory (DFT), self-consistent mean-field (SCMF) diffusion modeling, and an impurity trapping model to quantify the impact of defect trapping on effective vacancy mobility. Strong first-nearest-neighbor vacancy--Y binding, off-lattice vacancy configurations, and elevated rotational and dissociation barriers result in pronounced suppression of vacancy diffusivity, while Y$_2$O$_3$-associated sites treated as fixed traps in the model further reduce the mobile vacancy population. Ni--Mo (NM), Y-modified (NMY), and Y$_2$O$_3$-containing (NMYO) alloys were tested in molten fluoride salt at 750~\textdegree C. Corrosion penetration decreases systematically from NM to NMY to NMYO, consistent with the predicted reduction in vacancy-mediated transport. The same trend is observed in mass change, with net weight loss decreasing from 0.52\% for NM to 0.23\% for NMY and 0.09\% for NMYO. These results are consistent with an alloy-side transport contribution from defect trapping and reduced atomic transport. This study therefore suggests point defect engineering as a promising alloy design strategy for extreme nuclear environments.
}




\maketitle


\section{Introduction}\label{sec:introduction}
Molten salt reactors (MSRs) are among the most promising advanced nuclear energy systems due to their inherent safety characteristics, high operating temperatures, and efficient fuel utilization \cite{serp2014molten, roper2022molten,elsheikh2013safety}. However, the deployment of MSRs is fundamentally limited by the lack of structural materials capable of withstanding prolonged exposure to high-temperature molten fluoride salts \cite{busby2019technical,raiman2018aggregation}. In contrast to oxidation in air or aqueous corrosion, molten salt corrosion proceeds primarily through direct dissolution of alloying elements into the salt, as protective oxide films are thermodynamically unstable and readily dissolved under these conditions \cite{olson2009materials,olson2006evaluation,ignatiev2017corrosion,patel2017high,zhu2015high,ye2016high}. Therefore, many conventional structural alloys experience rapid degradation at the salt--alloy interface, particularly at elevated temperatures relevant to the MSR condition \cite{zheng2018corrosion,zheng2015high}.

Previous investigations to improve corrosion resistance in molten salt environments have focused largely on alloy chemistry optimization and thermodynamic stability considerations, such as minimizing elements with high fluoride solubility or selecting Ni-based alloys with favorable redox behavior \cite{olson2009materials,patel2017high,zheng2018corrosion,SRIDHARAN2013241,li2022effect}. While these approaches have led to improved corrosion performance, they primarily address equilibrium aspects of salt--alloy interactions and did not explicitly consider the kinetic processes governing corrosion. In practice, corrosion in molten salts is controlled not only by thermodynamic driving forces but also by the rate at which alloy atoms are transported from the bulk or near-surface matrix to the salt-facing interface, where dissolution reactions occur \cite{olson2009materials,SRIDHARAN2013241}. At high temperatures, this transport may include a contribution from vacancy-mediated diffusion.

From a kinetic viewpoint, suppressing atomic transport provides an alternative pathway to mitigating molten salt corrosion. By reducing vacancy mobility, the supply rate of dissolving species to the interface can be significantly lowered, even when thermodynamic driving forces remain unchanged. Despite its importance, diffusion control has received limited attention in molten salt corrosion research, particularly as a deliberate alloy design strategy. Minor solute additions and dispersed second-phase particles provide a means to influence vacancy behavior by acting as defect traps, thereby increasing the effective activation barrier for diffusion and reducing atomic mobility in the bulk and the near-surface regions of the alloy matrix.

Rare earth elements, such as yttrium (Y), are especially promising in this regard due to their large atomic size and strong interactions with point defects in Ni-based alloys. Experimental studies have demonstrated that minor Y additions can significantly improve corrosion resistance in molten fluoride salts, often attributed to the formation of stable Y-containing surface compounds \cite{li2022effect}. Theoretical and computational investigations further indicate that oversized solutes in alloys can strongly bind vacancies, causing reduced vacancy mobility and slower diffusion kinetics \cite{mansur1981effective,howard1965annealing,damask1960theory,flynn1964vacancy,le1978solute,ke2023microstructure,KE201930,ke2022first}. In parallel, oxide dispersions such as yttrium oxide introduce incoherent metal--oxide interfaces that are known to act as efficient sinks for point defects due to their reduced defect formation energies \cite{wang2023first,sun2018helium}. While these mechanisms have been widely studied for oxidation resistance and irradiation tolerance \cite{li2022effect,gan2004effect,kato1991effects}, their combined effect on diffusion-controlled molten salt corrosion has not been quantitatively established.

Recent experimental studies on Ni-based alloys provide direct evidence that atomic diffusion can be significantly suppressed through minor alloying additions. In particular, trace additions of yttrium have been shown to reduce effective atomic diffusion coefficients and slow diffusion-controlled microstructure evolution in Ni-based single-crystal superalloys at elevated temperatures \cite{wang2025unraveling}. Detailed kinetic analyses reveal a monotonic decrease in effective diffusion rates with increasing Y content, demonstrating that minor solute additions can significantly inhibit atomic mobility in Ni-based systems. While these studies were conducted in the context of high-temperature creep rather than corrosion, they show that diffusion in Ni alloys can be deliberately reduced under relevant thermal conditions. Notably, high-throughput and machine learning analyses of molten salt corrosion further identify the Ni diffusion coefficient as a dominant factor correlated with corrosion magnitude \cite{goh2024nobility}. Together, these observations motivate a corrosion mitigation strategy based on suppressing atomic mobility. For example, Y-induced defect trapping could reduce the transport of dissolving species to the salt--metal interface and thereby slow molten salt corrosion.

In this work, we address this gap by studying vacancy trapping from dilute yttrium solutes and dispersed yttrium oxide particles in Ni-based alloys. We quantify vacancy trapping by substitutional Y solutes and fixed $\text{Y}_2\text{O}_3$-associated trap sites in the model using first-principles calculations and defect-based diffusion models, predict the resulting reduction in effective vacancy mobility and its implications for vacancy-mediated atomic transport, and compare the predicted reduction in vacancy mobility with corrosion behavior measured in high-temperature molten fluoride salts. By evaluating a possible link between vacancy trapping, reduced atomic transport, and improved corrosion resistance, this study provides a general framework for alloy design strategies aimed at enhancing degradation resistance in extreme nuclear environments.

\section{Results}
\label{sec:results}
\subsection{Vacancy--yttrium binding}
\label{sect:vacancy_binding}

\begin{figure}
    \centering
    \includegraphics[scale=0.6]{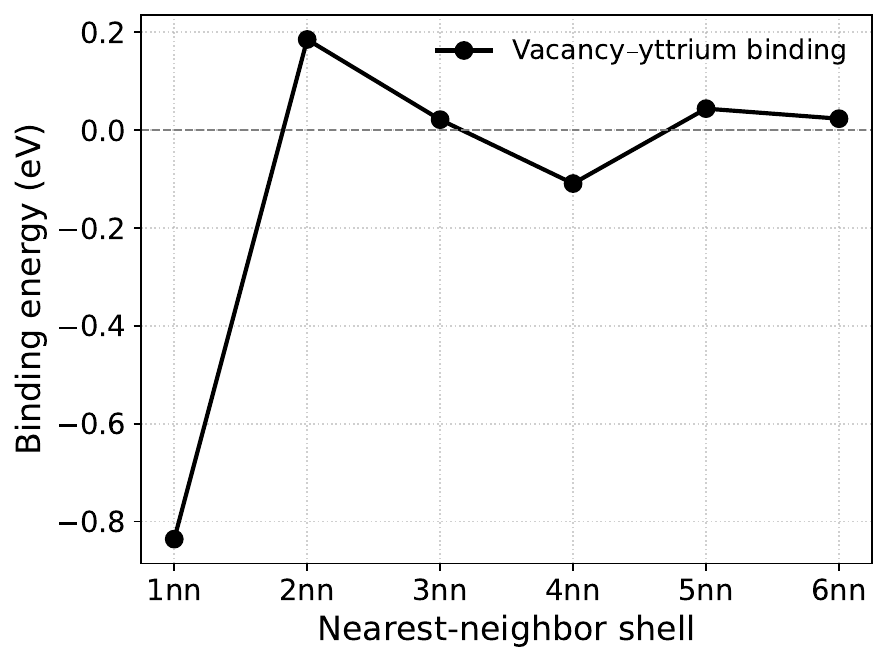}
    \caption{Plot showing the vacancy--yttrium binding energy as a function of nearest-neighbor (nn) separation in fcc Ni. Negative values indicate attractive interactions.}
    \label{fig:binding}
\end{figure}

We first examine vacancy--Y interactions at the atomic scale, which provide the basis for subsequent diffusion modeling and corrosion interpretation. Figure~\ref{fig:binding} shows the calculated vacancy--yttrium binding energy as a function of nearest-neighbor (nn) separation in face-centered cubic (fcc) Ni. A pronounced stabilization is obtained when the vacancy occupies the first-nearest-neighbor (1nn) configuration relative to Y, with an effective binding energy of approximately $-0.8$~eV. This value is significantly larger than the binding energies typically reported for conventional substitutional solutes in fcc Ni, indicating that Y acts as a strong vacancy trap.

As discussed in the Methods section, first-principles structural relaxations show that when a vacancy is placed at the 1nn position relative to Y, the equilibrium configuration becomes non-lattice-centered: the vacancy is no longer well described as a localized, lattice-centered defect adjacent to a substitutional solute. This strong relaxation indicates that the classical picture from the five-frequency model is not strictly applicable for the Y--vacancy exchange configuration, motivating the use of a more general transport framework in the following analysis.

Beyond the 1nn shell, the vacancy--Y interaction decays rapidly and becomes weak. The 2nn configuration is repulsive, the 3nn interaction is near-neutral, and a modest attractive interaction reappears at 4nn, while interactions at 5nn and 6nn are close to zero. This rapid attenuation indicates that vacancy trapping by Y is highly localized and dominated by the near-field environment. From a kinetic perspective, the deep 1nn stabilization implies long vacancy residence times near Y and a reduced population of mobile vacancies available for long-range diffusion, providing a microscopic basis for the diffusion suppression predicted by the transport calculations presented later.

\subsection{Migration energies}
\label{sect:migration}
\begin{figure}
    \centering
    \includegraphics[scale=0.69]{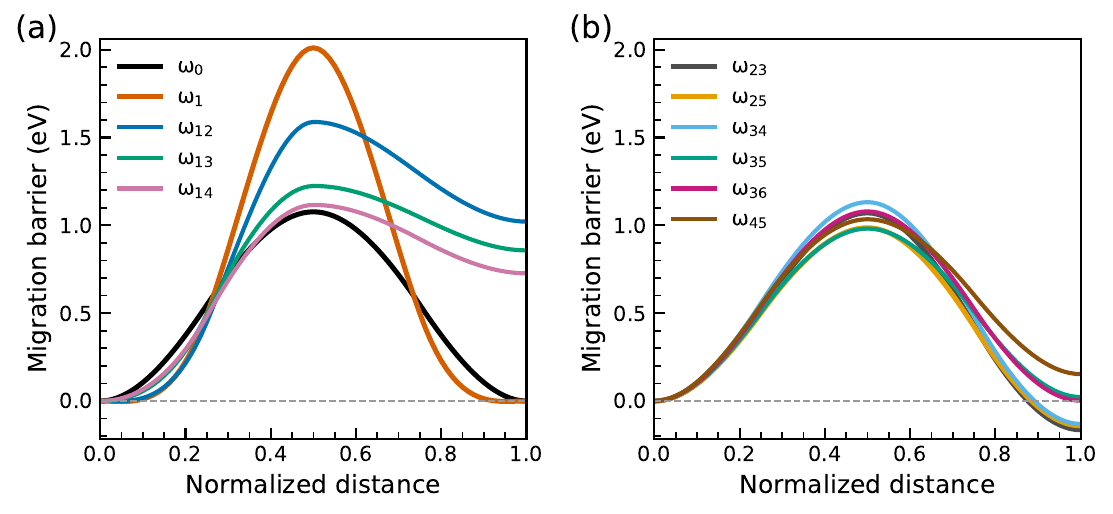}
    \caption{Migration energy profiles for vacancy jumps in dilute Y-containing fcc Ni calculated by CI-NEB within the multifrequency formalism. (a) Near-field hops involving the first-neighbor shell of Y, including $w_0$, $w_1$, and $w_{1n}$ ($n=2,3,4$). (b) Long-range hops beyond the near-field region ($w_{23}$, $w_{25}$, $w_{34}$, $w_{35}$, $w_{36}$, and $w_{45}$).}
    \label{fig:neb}
\end{figure}

Figure~\ref{fig:neb} shows the migration energy profiles obtained from climbing image nudged elastic band (CI-NEB) calculations for vacancy jumps in fcc Ni in the dilute Y limit, analyzed within the multifrequency framework for substitutional solute diffusion. The bulk-like hop $w_0$ shows a migration barrier of approximately 1.1~eV, corresponding to vacancy motion far from the solute and representative of Ni self diffusion. In contrast, the 1nn$\rightarrow$1nn rotational hop $w_1$, which describes vacancy reorientation within the first-neighbor shell of Y, shows a much higher barrier of $\sim 2.1$~eV, as shown in Figure~\ref{fig:neb}(a). This indicates that once the vacancy is captured at 1nn, its rotational mobility is strongly hindered. This reflects a stiff and highly distorted local solute--vacancy environment, consistent with the tendency of Y to induce a split vacancy configuration at the first neighbor site.

\begin{figure}
    \centering
    \includegraphics[scale=0.50]{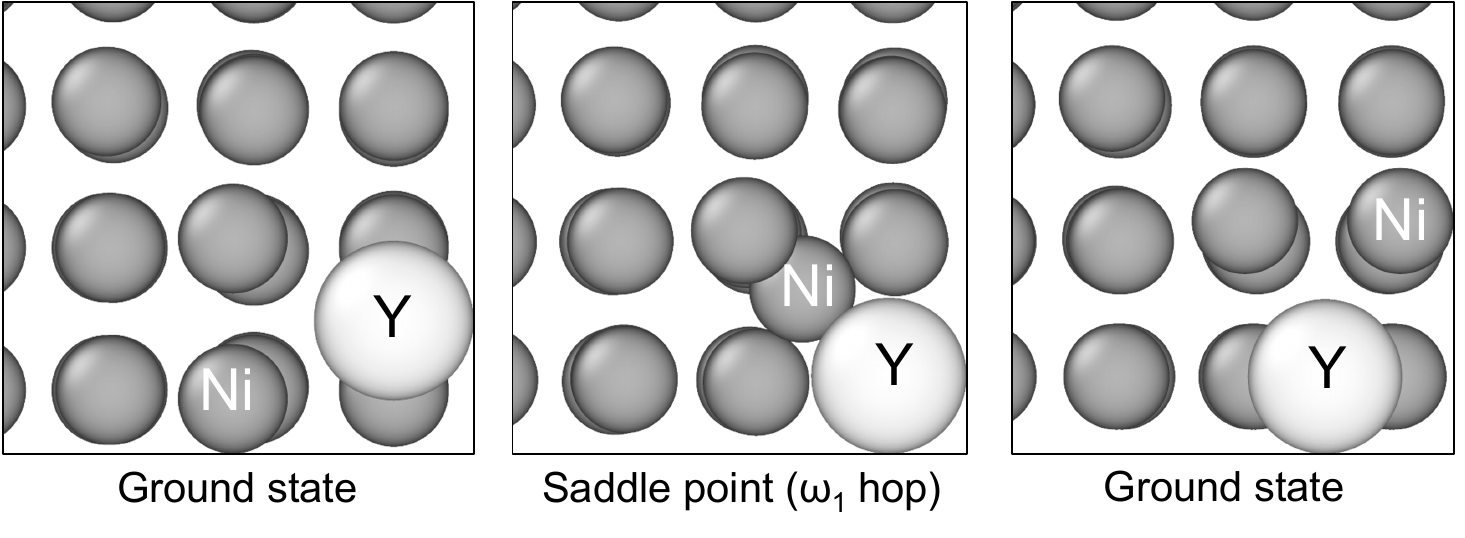}
    \caption{Atomic configuration of the 1nn$\rightarrow$1nn rotational ($w_1$) hop near Y in fcc Ni obtained from CI-NEB calculations. The vacancy forms a split vacancy configuration in the ground state (left). At the saddle point (center), cooperative rearrangement of surrounding Ni atoms is required, resulting in a high migration barrier. The final state (right) corresponds to an equivalent 1nn split vacancy configuration.}
    \label{fig:neb_w1}
\end{figure}

The unusually high barrier associated with the rotational hop $w_1$ can be understood from the atomic configurations of the split vacancy complex along the CI-NEB path (Figure~\ref{fig:neb_w1}). In the ground state, the vacancy is delocalized toward the oversized Y atom, which relaxes away from its ideal lattice position and forms a stabilized split vacancy configuration. Along the CI-NEB pathway, the saddle-point configuration shows that the migrating Ni atom must move into the neighboring 1nn site, reducing the Y off-lattice relaxation and driving the local environment toward a more constrained lattice-centered geometry. This constrained saddle-point geometry is the high-energy state along the CI-NEB path and therefore provides a structural basis for the large $w_1$ migration barrier relative to $w_0$. We interpret this connection from the relaxed atomic configurations and total energy profile, rather than from an atomic stress or atomic pressure measurement.

Figure~\ref{fig:neb}(a) further shows that the dissociation hops $w_{21}$, $w_{31}$, and $w_{41}$ exhibit migration barriers that are clearly elevated relative to $w_0$ and the association hops. This implies kinetically hindered escape from the bound configuration and is consistent with the strong 1nn vacancy binding energy of $\sim 0.8$~eV shown in Figure~\ref{fig:binding}. The elevated dissociation barriers and energetic stabilization of the 1nn configuration indicate the formation of a stable vacancy--Y complex. In the dilute limit, such strong binding produces a pronounced trapping effect near the solute and significantly increases the vacancy residence time within the first coordination shell.

In contrast, hops to more distant shells shown in Figure~\ref{fig:neb}(b) rapidly recover toward the bulk barrier $w_0$. The migration energy profiles for these longer-range jumps collapse closely onto $w_0$ with only minor variation. This suggests that the perturbation induced by Y is highly localized and confined within the first few coordination shells. Therefore, within the multifrequency formalism, far-field jump frequencies may be approximated by $w_0$ without significant loss of accuracy, thereby reducing the number of independent kinetic parameters required to describe vacancy-mediated transport.

\subsection{Effect of Y on vacancy mobility}
\label{sect:vac_mob_Y}
\begin{figure}
    \centering
    \includegraphics[scale=0.65]{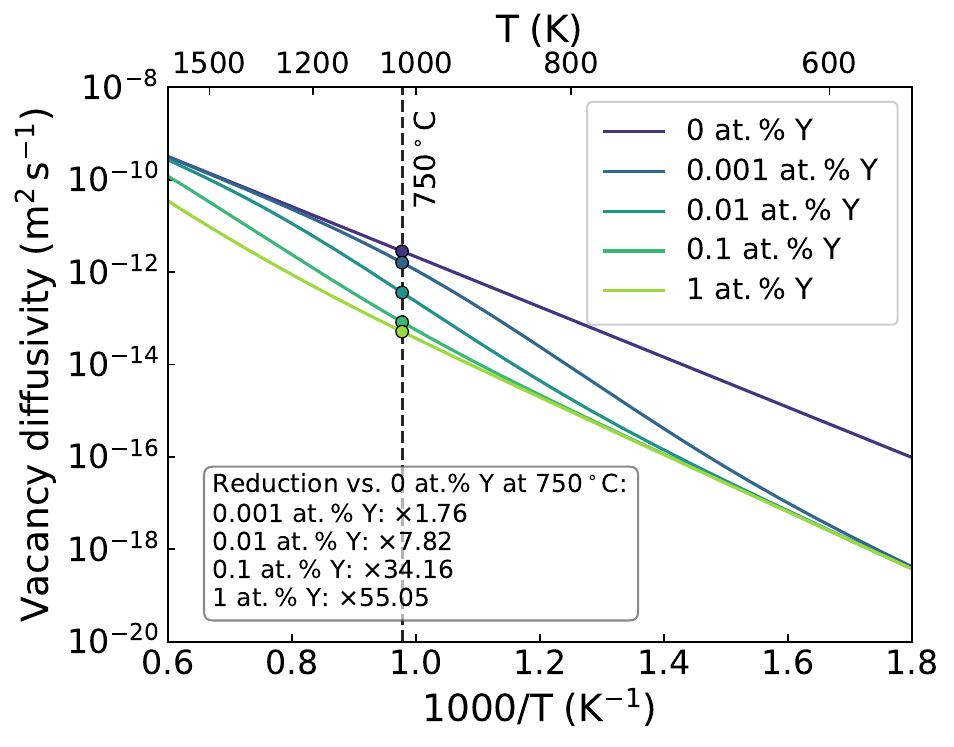}
    \caption{Vacancy diffusion coefficient plotted as a function of inverse temperature (1000/$T$) for fcc Ni with 0, 0.001, 0.01, 0.1, and 1 at.\% Y. The upper axis shows the corresponding temperature in Kelvin. The marked values at 750~\textdegree C highlight the concentration dependent reduction in vacancy mobility compared to pure Ni.}
    \label{fig:D_V}
\end{figure}

The density functional theory (DFT) results presented above are used as input parameters in the self-consistent mean-field (SCMF) framework to evaluate transport coefficients. The calculated vacancy diffusion coefficient for Ni containing 0, 0.001, 0.01, 0.1, and 1 at.\% Y is shown in Figure~\ref{fig:D_V} as a function of inverse temperature.

In the absence of Y (0 at.\%), the vacancy diffusivity follows a linear Arrhenius behavior, suggesting thermally activated bulk vacancy migration governed by the intrinsic migration barrier of Ni. Upon the introduction of dilute Y, the vacancy mobility is progressively reduced across the entire temperature range. Even at very dilute concentrations (0.001 at.\%), a measurable decrease in diffusivity is observed, while at higher concentrations (0.1--1 at.\%) the reduction becomes significant, particularly at lower temperatures.

The reduction in vacancy mobility originates from the strong short-range vacancy--Y binding and the elevated migration barriers near the solute, as discussed above. The large 1nn binding energy promotes vacancy trapping in the vicinity of Y atoms, effectively decreasing the fraction of mobile (free) vacancies. In addition, the suppressed rotational frequency $w_1$ and the elevated dissociation barriers hinder local reorientation and escape from the bound configuration. Within the SCMF formalism, these effects enter through modified jump frequencies and correlation factors, resulting in a reduced effective transport coefficient.

The influence of Y becomes clearer as the system enters the vacancy trapping regime on cooling, but the curves for alloys containing Y approach a common mobility limit controlled by trapped complexes at the lowest temperatures. This behavior is expected since vacancy trapping is thermodynamically favored at low temperature, where the Boltzmann factor associated with the 1nn binding energy enhances the equilibrium concentration of vacancy--Y complexes. Once most vacancies are trapped, the remaining long-range motion is controlled mainly by the mobility and escape kinetics of strongly bound vacancy--Y complexes. Therefore, further increasing the Y concentration produces a smaller additional change at the lowest temperatures. As temperature increases, thermal activation facilitates dissociation from the bound state, partially restoring bulk-like vacancy mobility and reducing the concentration dependent deviation from the pure Ni case.

It can also be seen in Figure~\ref{fig:D_V} that the vacancy diffusivity in Y-containing alloys deviates from ideal Arrhenius behavior, particularly in the temperature range where the relative contributions of free vacancies and vacancy--Y complexes change most rapidly. This non-Arrhenius feature is caused by the coexistence and competition between two distinct transport channels: migration by free vacancies and migration by vacancy--Y complexes, similar to the cases reported in \cite{KE201930,ke2022first}. At low temperatures, vacancy diffusion is dominated by trapped complexes whose mobility is governed by the higher rotational and dissociation barriers. At higher temperatures, dissociation becomes more probable and transport increasingly proceeds through free vacancies with the bulk-like barrier $w_0$. Because these two mechanisms show different effective activation energies, the temperature-dependent transition in their relative contributions produces curvature in the Arrhenius plot. This corresponds to a temperature-dependent redistribution between bound and unbound vacancy populations, resulting in activation energy that varies with temperature.

Overall, Figure~\ref{fig:D_V} demonstrates that even extremely small additions of Y can lead to a measurable suppression of vacancy mobility in fcc Ni due to strong short-range trapping and modified local migration kinetics. At 750~\textdegree C, the effective vacancy diffusivity in the presence of only 0.001\,at.\% Y is reduced by a factor of approximately 1.8 relative to pure Ni, with this suppression increasing to factors of 7.8, 34.2, and 55.1 for 0.01\,at.\%, 0.1\,at.\%, and 1\,at.\% Y, respectively. The pronounced impact at very low solute content indicates that vacancy trapping is highly effective even at trace levels, which is of practical significance given that Y is typically added in dilute concentrations in structural Ni-based alloys.

Note that the effect of Y on vacancy mobility shown in Fig.~\ref{fig:D_V} is qualitatively opposite to the behavior previously reported for phosphorus in Ni-P alloys \cite{KE201930}. In the P case, strong vacancy binding ($E_{vb} \approx 0.3$~eV) was accompanied by significantly reduced rotation ($w_1$) and exchange ($w_2$) barriers relative to the bulk hop ($w_0$). As a result, vacancy--solute pairs were highly mobile and diffusion proceeded efficiently via the Johnson mechanism \cite{johnson1939note}, resulting in enhanced solvent diffusion and increased effective vacancy mobility.

\subsection{Effect of Y$_2$O$_3$ on vacancy mobility}
\label{sect:vac_mob_Y2O3}

\begin{figure}
    \centering
    \includegraphics[scale=0.65]{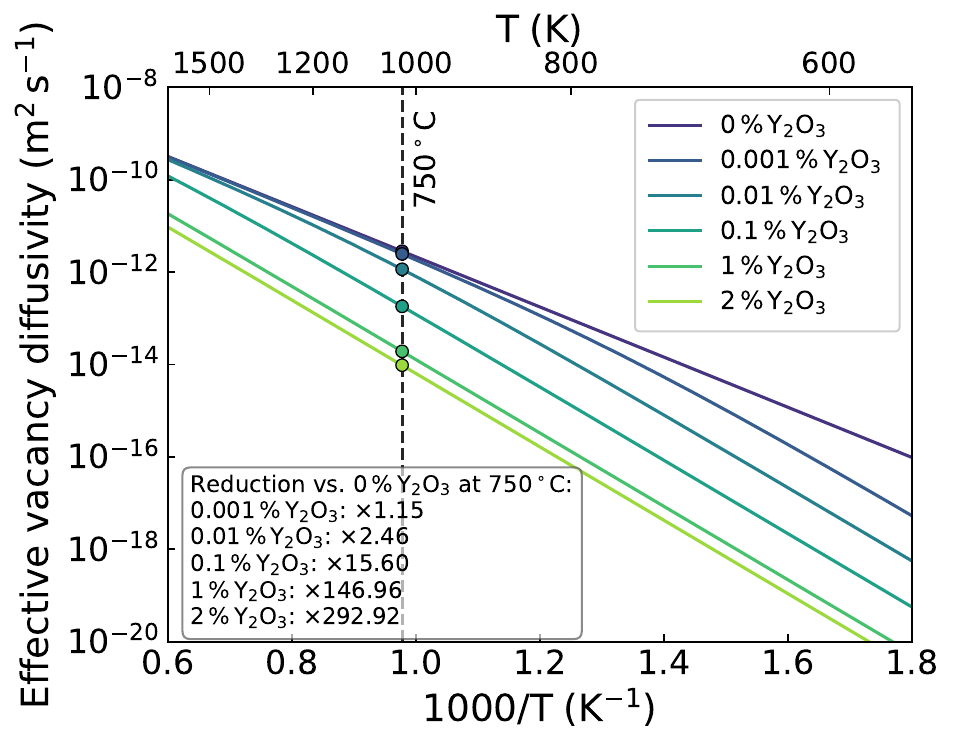}
    \caption{Effective vacancy diffusion coefficient as a function of inverse temperature (1000/$T$) for fcc Ni containing 0, 0.001, 0.01, 0.1, 1, and 2 vol.\% Y$_2$O$_3$, calculated using the impurity trapping model with a vacancy--trap binding energy of 0.5 eV.}
    \label{fig:D_V_Y2O3}
\end{figure}

The influence of Y$_2$O$_3$ particles on vacancy transport was evaluated using the impurity trapping model described in the Methods section. In this framework, sites associated with Y$_2$O$_3$ particles are treated as fixed vacancy traps characterized by a vacancy--trap binding energy of 0.5~eV. The effective vacancy diffusivity was calculated as a function of temperature for Ni containing 0, 0.001, 0.01, 0.1, 1, and 2~vol.\% Y$_2$O$_3$, as shown in Figure~\ref{fig:D_V_Y2O3}. Compared to the substitutional Y case, Y$_2$O$_3$ produces a stronger suppression of vacancy mobility across the entire temperature range.

Within the model, vacancies captured at fixed oxide-associated trap sites are removed from the mobile population until thermally released. In contrast to the SCMF description for substitutional Y, where bound vacancy--solute complexes may still contribute to transport, Y$_2$O$_3$ trapping reduces the effective concentration of free vacancies. Therefore, vacancy diffusivity decreases monotonically with increasing oxide content. At 750~\textdegree C, the effective vacancy diffusivity is reduced by factors of approximately 1.15, 2.46, 15.60, 146.96, and 292.92 for 0.001, 0.01, 0.1, 1, and 2 vol.\% Y$_2$O$_3$, respectively, relative to pure Ni.

Note that the vacancy--Y$_2$O$_3$ binding energy was assumed to be 0.5~eV in the impurity trapping model. This choice is guided by first-principles studies of metal/oxide interfaces, which have shown reduced vacancy formation energies near such interfaces. For example, Kumar \textit{et al.} reported that the Ni vacancy formation energy at a coherent Ni/Al$_2$O$_3$ interface decreases from approximately 1.25~eV in bulk Ni to 0.85~eV near the interface, corresponding to an effective stabilization of about 0.4~eV \cite{kumar2018anomalous}. In contrast, Peng and Lai found significantly larger vacancy stabilization at a coherent $\alpha$-Fe/Y$_2$O$_3$ interface, including negative vacancy formation energies in specific interfacial layers Ref.~\cite{peng2015interaction}. These results indicate that oxide interfaces can strongly attract and stabilize vacancies. Nevertheless, the magnitude of the interaction depends sensitively on interface coherency, local bonding environment, lattice misfit, and chemical composition. The Fe/Y$_2$O$_3$ system considered in Ref.~\cite{peng2015interaction} corresponds to a highly coherent cube-on-cube interface under specific strain conditions, which may overestimate vacancy stabilization compared to the incoherent or semi-coherent Y$_2$O$_3$/Ni interfaces that are typically present in Ni alloys strengthened by oxide dispersion. Therefore, the selected value of 0.5~eV represents a moderate and physically reasonable estimate within the range suggested by the previous study \cite{kumar2018anomalous}, while avoiding the potentially extreme stabilization predicted for ideal coherent interfaces \cite{peng2015interaction}. Note that the accurate vacancy binding energy at an incoherent Y$_2$O$_3$/Ni interface remains uncertain and would require systematic atomistic modeling of realistic interface structures.

From a materials design perspective, the ability to significantly modify vacancy transport with minimal Y or Y$_2$O$_3$ addition is potentially attractive for applications where vacancy-mediated processes (such as creep, precipitation, and microstructure evolution) contribute to performance or degradation. While Y is a rare-earth element, Y can be used economically at dilute levels in many alloy systems. The observed reductions in vacancy diffusivity suggest that targeted alloying Y or Y$_2$O$_3$ can achieve significant changes in transport behavior without requiring high solute loadings, thereby offering a favorable combination of performance modification and alloy cost. Together, substitutional Y and oxide dispersion represent complementary vacancy trapping mechanisms operating at atomic and mesoscale levels. Their combined influence is expected to produce cumulative suppression of vacancy-mediated transport.

\subsection{Corrosion response of NM, NMY, and NMYO alloys}
\label{sec:corrosion_exp}
\begin{figure}
    \centering
    \includegraphics[scale=0.41]{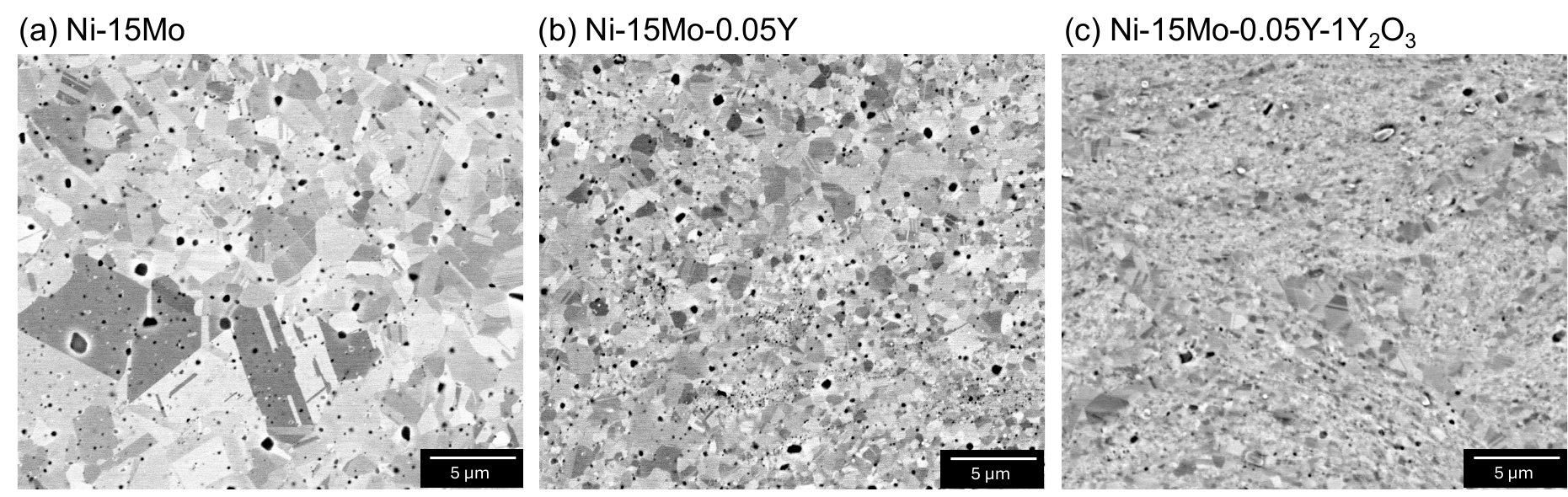}
    \caption{Backscattered electron (BSE) micrographs of the as-fabricated alloys before molten salt exposure: (a) Ni--15Mo (NM), (b) Ni--15Mo--0.05Y (NMY), and (c) Ni--15Mo--0.05Y--1Y$_2$O$_3$ (NMYO). Based on the characterization reported by Zhang \textit{et al.} \cite{zhang2026ods}, the dark contrast particles include MoO$_2$ in NM/NMY and Y-enriched or Y$_2$O$_3$ particles in NMYO.}
    \label{fig:as_fabricated}
\end{figure}

Having established the suppression of vacancy mobility through both solute trapping and oxide dispersion, we then examine whether these predicted reductions translate into measurable corrosion resistance. Figure~\ref{fig:as_fabricated} shows the backscattered electron (BSE) microstructures of the as-fabricated Ni--15Mo (NM), Ni--15Mo--0.05Y (NMY), and Ni--15Mo--0.05Y--1Y$_2$O$_3$ (NMYO) alloys before molten salt exposure. All three compositions show dense and consolidated microstructures, confirming the effectiveness of the electric field assisted sintering (EFAS) processing (refer to the Methods section). The NM alloy shows a relatively equiaxed grain structure with clearly defined grain boundaries. The addition of 0.05 wt.\% Y in NMY refines the grain structure, with the grain size measured by electron backscatter diffraction (EBSD) decreasing from $1.37 \pm 0.91~\mu$m for NM to $0.70 \pm 0.33~\mu$m for NMY (Table~\ref{tab:grain_size_summary}) \cite{zhang2026ods}. In the NMYO alloy, the alloy containing Y$_2$O$_3$ remains fine-grained, with an EBSD grain size of $0.50 \pm 0.36~\mu$m (Table~\ref{tab:grain_size_summary}) \cite{zhang2026ods}. The corresponding EBSD inverse pole figure (IPF) maps and grain size distributions are provided in Supplementary Fig.~S1. Detailed EBSD maps, grain size distributions, phase identification, and as-fabricated transmission electron microscopy/high-resolution transmission electron microscopy (TEM/HRTEM) characterization are reported by Zhang \textit{et al.} \cite{zhang2026ods}. That study identified MoO$_2$ particles in NM/NMY. These observations clarify that Mo is not completely dissolved in the matrix. Rather, the alloys are best described as a predominantly fcc Ni--Mo matrix containing oxide secondary phases related to processing. The dark contrast particles in Fig.~\ref{fig:as_fabricated} therefore include MoO$_2$ in NM/NMY and Y-enriched or Y$_2$O$_3$ particles in NMYO. A carbon map from an additional as-fabricated NM field is provided in Supplementary Fig.~S3 and shows no obvious localized carbon enrichment at the dark particles in that field. Because carbon was not mapped in the original EDS field, a minor contribution from carbon-bearing particles cannot be excluded. The MoO$_2$ particles are likely related to residual oxygen associated with powder processing and EFAS consolidation. The same characterization further identified regions enriched in Y and Y$_2$O$_3$ in NMY and discrete crystalline Y$_2$O$_3$ nanoparticles in NMYO by scanning electron microscopy with energy dispersive X-ray spectroscopy (SEM/EDS), X-ray diffraction (XRD), scanning transmission electron microscopy with energy dispersive X-ray spectroscopy (STEM-EDS), and HRTEM analyses \cite{zhang2026ods}. Grain refinement can influence molten salt corrosion because grain boundaries may provide fast diffusion and attack pathways. However, NMY and NMYO have higher, not lower, grain boundary densities than NM. Therefore, the reduced corrosion depth in the alloys containing Y and Y$_2$O$_3$ cannot be explained by a simple reduction in grain boundary area alone. The absence of pre-exposure damage across all three alloys allows us to evaluate the subsequent corrosion behavior in relation to the calculated defect transport trends while acknowledging microstructural contributions.

\begin{table}[!htbp]
\centering
\caption{EBSD grain sizes for NM, NMY, and NMYO \cite{zhang2026ods}.}
\label{tab:grain_size_summary}
{\begin{tabular}{lc}
\toprule
Material & Grain size (\textmu m) \\
\midrule
NM & $1.37 \pm 0.91$ \\
NMY & $0.70 \pm 0.33$ \\
NMYO & $0.50 \pm 0.36$ \\
\bottomrule
\end{tabular}
}
\end{table}

\begin{figure}
    \centering
    \includegraphics[scale=0.45]{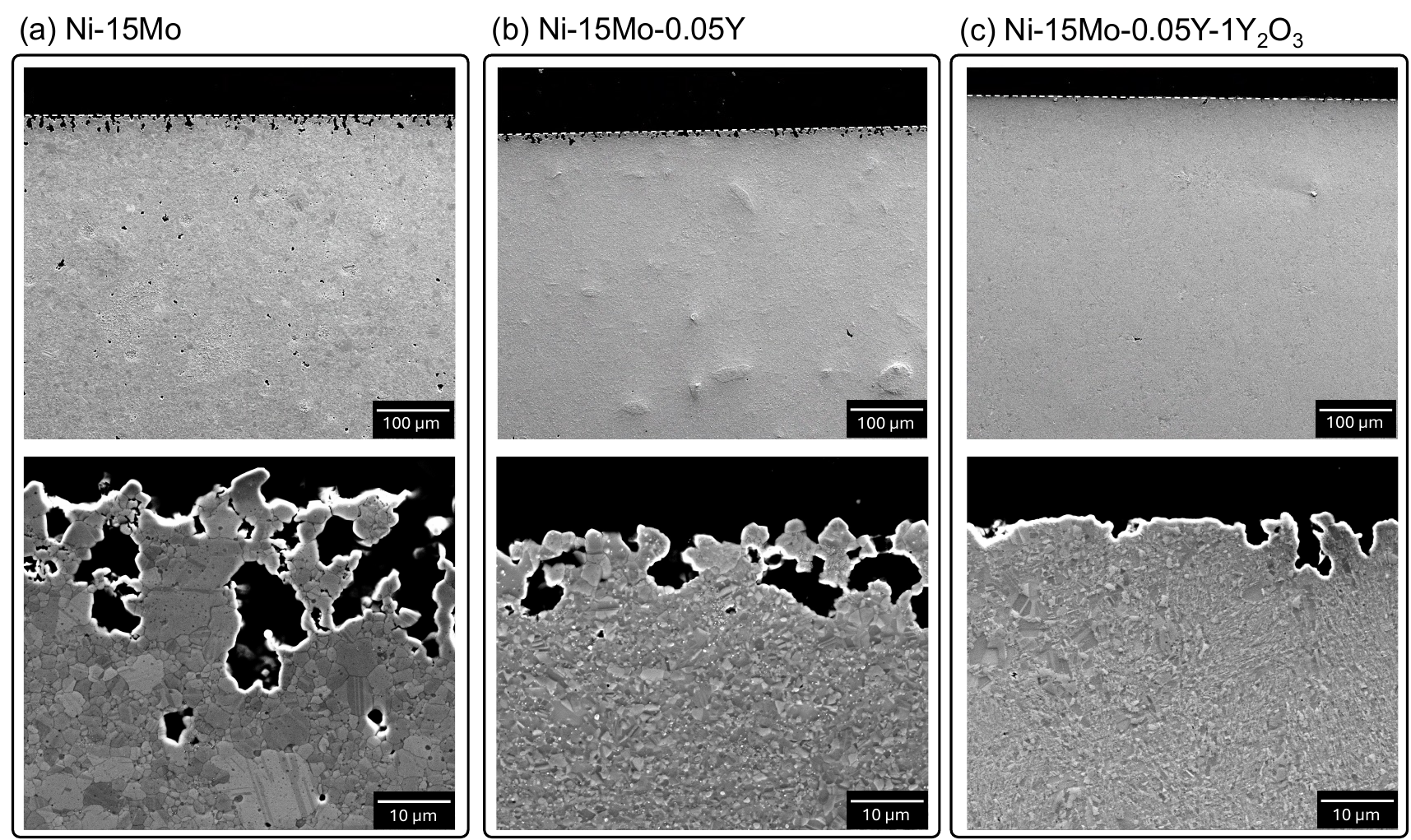}
    \caption{Cross-sectional SEM images of (a) Ni--15Mo (NM), (b) Ni--15Mo--0.05Y (NMY), and (c) Ni--15Mo--0.05Y--1Y$_2$O$_3$ (NMYO) alloys after molten salt exposure. The top row shows 500X views of the overall attacked region, while the bottom row shows 5000X near-surface images. White dashed lines mark the visible remaining exposed surface used as the SEM depth reference, not a reconstructed pre-exposure salt--metal interface.}
    \label{fig:corrosion_SEM}
\end{figure}

To assess whether suppression of defect mobility translates into measurable corrosion resistance, the corrosion response of NM, NMY, and NMYO alloys was evaluated. Figure \ref{fig:corrosion_SEM} presents representative cross-sectional SEM images of the NM, NMY, and NMYO alloys after exposure to molten FLiNaK salt at 750~\textdegree C for 500~h. The extent of corrosion and penetration depth decrease from NM to NMY to NMYO. The NM alloy shows the most severe degradation, with an estimated corrosion penetration depth of $\sim$25--30~\textmu m. The maximum penetration ranges were estimated from cross-sectional SEM images as the distance from the visible remaining exposed surface to the deepest continuous region of corrosion porosity, surface recession, or near-surface attack. The white dashed lines in Figure~\ref{fig:corrosion_SEM} are visual guides added to show this practical SEM reference surface. Because the original salt--metal interface was not independently marked before exposure, these values are reported as approximate ranges. Extensive near-surface porosity and interconnected attack features form a corrosion front that penetrates deeply into the substrate. The NM corrosion morphology shows some similarities to dealloying morphologies reported for molten salt dealloying and Ni--Cr corrosion, including porous and interconnected attack features \cite{liu2021formation,liu2023temperature}. This comparison is made only at the level of morphology. Because the present alloys are Ni--Mo model alloys without intentional Cr addition, we do not interpret the attack as the conventional selective dissolution of Cr used to explain many Ni--Cr alloys. A more complete picture of the local corrosion pathway would require more comprehensive post-corrosion elemental mapping and phase characterization.

In contrast, the NMY alloy shows a reduced corrosion depth of $\sim$10--15~\textmu m and a more discontinuous corrosion front. Although surface roughening and localized porosity remain visible, the porous features are smaller and less interconnected compared to NM. These observations indicate less extensive corrosion damage than in NM and are consistent with, but do not isolate, a possible contribution from Y-induced vacancy trapping.

The NMYO alloy demonstrates the best corrosion resistance among the three systems, with only shallow surface recession and a limited penetration depth of $\sim$5--7~\textmu m. Subsurface porosity is also minimal, and the grain structure remains largely intact. The corrosion front is less developed and more uniform compared to that of NM and NMY (Figure \ref{fig:corrosion_SEM} (a) and (b), respectively). These features show that the corrosion damage is less extensive, although the cross-sectional morphology alone does not identify the controlling transport or reaction step.

\begin{figure}[!htbp]
    \centering
    \includegraphics[scale=0.70]{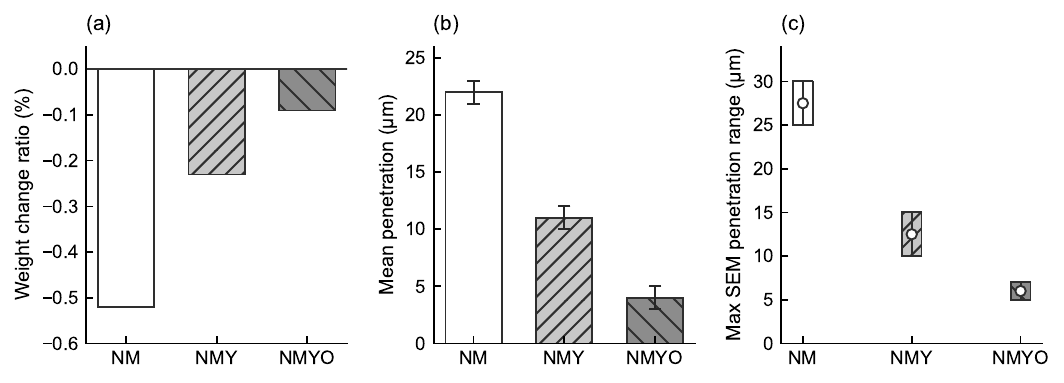}
    \caption{Corrosion response metrics for NM, NMY, and NMYO after FLiNaK exposure at 750~\textdegree C for 500~h: (a) weight change ratio, (b) mean penetration from two specimens, and (c) maximum SEM penetration range. Error bars in (b) show one standard deviation, and range bars in (c) show the approximate SEM penetration ranges.}
    \label{fig:corrosion_metrics_summary}
\end{figure}

The observed trend (NM $>$ NMY $>$ NMYO in both corrosion extent and maximum penetration depth) can be compared with the modeling results presented in the Results section, where strong vacancy binding by substitutional Y and additional fixed vacancy trapping at Y$_2$O$_3$ interfaces lead to reduced effective defect mobility. Among the three alloys, NMYO includes both solute-induced vacancy trapping and oxide-associated vacancy trapping, and it also shows the lowest measured corrosion damage. The same ranking is supported by the weight change ratio, mean penetration from two specimens, and maximum SEM penetration range summarized in Figure~\ref{fig:corrosion_metrics_summary} and Table~\ref{tab:corrosion_metrics}. The mean penetration depths are $22 \pm 1$, $11 \pm 1$, and $4 \pm 1$~\textmu m for NM, NMY, and NMYO, respectively. These uncertainties represent the variation between the average penetration depths from the two specimens. The local variation caused by surface roughness and nonuniform attack is larger and is captured by the repeated 500X line measurements reported in Table~\ref{tab:corrosion_line_measurements}. When normalized to NM, the weight loss magnitude and mean penetration decrease to 0.44 and 0.50 for NMY, and to 0.17 and 0.18 for NMYO, respectively. This agreement between the mass loss and SEM penetration metrics supports the interpretation that the lower penetration in NMYO is accompanied by lower overall material loss, rather than simply a flatter corrosion morphology with comparable mass removal. Additional cross-sectional SEM images from the second specimen are provided in Supplementary Fig.~S2. These results are consistent with a contribution from reduced defect transport to the improved corrosion resistance of NMYO.

\begin{table}[!htbp]
\centering
\caption{Corrosion response metrics after FLiNaK exposure at 750~\textdegree C for 500~h. Mean penetration values are calculated from the two specimen average depths. The maximum SEM penetration range describes the deepest continuous local attack observed in cross section.}
\label{tab:corrosion_metrics}
{\small
\setlength{\tabcolsep}{4pt}
\renewcommand{\arraystretch}{1.15}
\begin{tabular}{@{}lccc@{}}
\toprule
Material & Wt. change & \shortstack{Mean\\penetration} & \shortstack{Max SEM\\penetration\\range} \\
 & (\%) & (\textmu m) & (\textmu m) \\
\midrule
NM & $-0.52$ & $22 \pm 1$ & 25--30 \\
NMY & $-0.23$ & $11 \pm 1$ & 10--15 \\
NMYO & $-0.09$ & $4 \pm 1$ & 5--7 \\
\bottomrule
\end{tabular}
}
\end{table}

\begin{table}[!htbp]
\centering
\caption{500X SEM penetration measurements from two specimens of each composition after FLiNaK exposure at 750~\textdegree C for 500~h. Values are reported as mean $\pm$ one standard deviation. The column for both specimens is calculated from all individual 500X line measurements from sample 1 and sample 2.}
\label{tab:corrosion_line_measurements}
{\small
\setlength{\tabcolsep}{4pt}
\renewcommand{\arraystretch}{1.15}
\begin{tabular}{@{}lcccc@{}}
\toprule
Material & \shortstack{Sample 1\\line penetration\\(\textmu m)} & \shortstack{Sample 2\\line penetration\\(\textmu m)} & \shortstack{Line measurements\\from both\\specimens\\(\textmu m)} & \shortstack{Mean penetration\\of two specimens\\(\textmu m)} \\
\midrule
NM & $21 \pm 5$ & $23 \pm 9$ & $22 \pm 7$ & $22 \pm 1$ \\
NMY & $10 \pm 2$ & $11 \pm 5$ & $11 \pm 4$ & $11 \pm 1$ \\
NMYO & $3 \pm 1$ & $4 \pm 2$ & $3 \pm 2$ & $4 \pm 1$ \\
\bottomrule
\end{tabular}
}
\end{table}

After exposure, FLiNaK salts were analyzed by inductively coupled plasma mass spectrometry (ICP-MS), and the measured Ni, Mo, and Y concentrations are summarized in Table~\ref{tab:post_test_icpms}. Among the metallic species reported here, Mo shows the highest concentrations in the salt, with 231 parts per million (ppm) for NM, 100 ppm for NMY, and 75 ppm for NMYO, while Ni remains near 1--2 ppm. Thus, Mo has the highest concentration among the metallic species measured after exposure of these Ni--Mo model alloys without intentional Cr addition. These measurements show that Mo is present in the salt after exposure, but they do not determine its chemical form, the interfacial reaction sequence, or the rate-controlling process.

The present ICP-MS analysis quantified selected metallic species only. Sulfur and oxygen were not directly quantified in the starting or post-exposure salts. Oxygen-containing and sulfur-containing impurities can alter corrosion reactions and morphology in FLiNaK \cite{liu2023oxide,xu2021sulfate}. Therefore, the present data do not support attributing the measured Mo concentrations to a specific oxygen-driven or sulfur-driven reaction pathway. Y is reported in Table~\ref{tab:post_test_icpms} for completeness, with measured concentrations of 0, 0.9, and 10 ppm for NM, NMY, and NMYO, respectively. The higher Y concentration measured in the salt from the NMYO test suggests possible interaction or partial dissolution of the added Y$_2$O$_3$ dispersion. However, ICP-MS alone does not identify the source or chemical state of Y. The cross-sectional morphology of NM has similarities to dealloying morphologies reported for molten salt corrosion of Ni--Cr alloys \cite{liu2021formation,liu2023temperature}, but this morphological comparison does not show that the same dissolution chemistry applies.

\begin{table}[!htbp]
\centering
\caption{ICP-MS concentrations of selected metallic species in FLiNaK after 500 h exposure at 750~\textdegree C.}
\label{tab:post_test_icpms}
{\begin{tabular*}{0.82\linewidth}{@{\extracolsep{\fill}}lccc@{}}
\toprule
Specimen & Ni (ppm) & Y (ppm) & Mo (ppm) \\
\midrule
NM & 1.8 & 0 & 231 \\
NMY & 1.0 & 0.9 & 100 \\
NMYO & 1.3 & 10 & 75 \\
\bottomrule
\end{tabular*}
}
\end{table}

With the experimental corrosion trend summarized, we next compare the tested Y and Y$_2$O$_3$ levels with the calculated vacancy mobility reductions. Specifically, 0.05~wt.\% Y and 1~wt.\% Y$_2$O$_3$ correspond to approximately 0.035~at.\% Y and 1.78~vol.\% Y$_2$O$_3$, respectively. Based on the calculated vacancy diffusivities ($D_\text{V}$) and effective vacancy diffusivities ($D^{\mathrm{eff}}_{\mathrm{V}}$) described in the Methods section (see also Figures~\ref{fig:D_V} and \ref{fig:D_V_Y2O3}), the reduction in vacancy mobility at 750~\textdegree C is approximately 17 times for 0.035~at.\% Y and 261 times for 1.78~vol.\% Y$_2$O$_3$. Thus, the suppression associated with 1~wt.\% Y$_2$O$_3$ is roughly an order of magnitude greater than that produced by 0.05~wt.\% Y alone, consistent with the experimental observation in Figure~\ref{fig:corrosion_SEM} that the corrosion resistance of NMYO exceeds that of NMY.

Corrosion in this system involves coupled interfacial reactions and alloy-side transport. The present model addresses only alloy-side vacancy-mediated transport, and the current data do not show that this contribution controls the overall corrosion rate.

\FloatBarrier

\section{Discussion}
\label{sec:discussion}

Table~\ref{table:DFT_data_binding} compares vacancy--solute binding energies reported for selected solutes in fcc Ni and their size mismatch relative to Ni, quantified using Goldschmidt metallic radii. For moderately oversized refractory and transition-metal solutes (Nb, Mo, Ta, W, and Re), the binding energies are relatively small (generally within $\pm 0.1$~eV) and often exhibit an oscillatory, diminishing dependence on separation distance. In several cases, the strongest attractive interaction occurs at 4nn rather than 1nn, consistent with predominantly elastic size-misfit interactions in which the vacancy partially relieves local strain while remaining a well-defined lattice defect.

In contrast, Y exhibits qualitatively different behavior. Because of its much larger atomic radius (approximately 40--45\% larger than Ni), Y shows a strong and highly localized 1nn attraction ($\sim$--0.84~eV). This degree of oversizing is consistent with the type-III solute category proposed by Naghavi \textit{et al.}~\cite{NAGHAVI2017467}, where extreme size mismatch can induce pronounced local relaxation, strong vacancy trapping, and non-classical diffusion configurations. On the basis of size mismatch alone, other large solutes such as Zr, Hf, and Sc (with mismatches on the order of 28--32\% relative to Ni) may also be candidates for enhanced vacancy trapping. However, quantitative assessment of binding energies and migration barriers would be required to determine whether they exhibit behavior comparable to Y in fcc Ni.

It is also worth noting that the 1nn vacancy--Y binding energy reported here is lower than the value of 1.26~eV reported in Ref.~\cite{HARGATHER2018126}, where a 32-atom ($2\times2\times2$) supercell was employed. In such a small periodic cell, the periodic images of Y are separated by only two lattice spacings, which can introduce artificial Y--Y interactions and overlapping elastic strain fields, particularly for highly oversized solutes. This constraint can hinder elastic relaxation and bias the calculated energetics, emphasizing the importance of supercell size convergence when treating extreme size-mismatch solutes.

\begin{table*}[t]
\small
\caption{DFT-calculated vacancy binding energies of solute atoms from the first- to sixth-nearest-neighbor (1nn--6nn) positions in fcc Ni. Binding energies are given in eV. Negative values indicate attractive interaction under the present sign convention. Goldschmidt (metallic) atomic radii and corresponding size mismatch relative to Ni (124 pm) are also shown.}
\centering
\setlength{\tabcolsep}{4pt}
\resizebox{\textwidth}{!}{%
\begin{tabular}{lccccccccc}
\hline
Solute & Radius (pm) & $\Delta r/r_{\text{Ni}}$ (\%) & 1nn & 2nn & 3nn & 4nn & 5nn & 6nn & Reference \\
\hline
Y  & 180 & +45 & -0.84 &  0.19 &  0.02 & -0.11 &  0.04 &  0.02 & This work \\
Nb & 146 & +18 & -0.13 &  0.06 &  0.03 & -0.07 &  --   &  --   & \cite{tucker2014solute} \\
Mo & 139 & +12 & -0.01 & -0.02 &  0.03 & -0.07 &  0.01 & -0.01 & \cite{shousha2024vacancy} \\
Ta & 146 & +18 & -0.08 &  0.03 &  0.03 & -0.09 &  0.01 & -0.01 & \cite{shousha2024vacancy} \\
W  & 139 & +12 &  0.02 & -0.02 &  0.04 & -0.06 &  0.01 & -0.01 & \cite{shousha2024vacancy} \\
Re & 137 & +10 &  0.04 & -0.07 &  0.03 & -0.05 &  0.01 & -0.02 & \cite{shousha2024vacancy} \\
\hline
\end{tabular}
}
\label{table:DFT_data_binding}
\end{table*}

Finally, the relevance of oversized solute vacancy trapping extends beyond fundamental defect energetics to irradiation performance. Oversized additions have been reported to mitigate radiation-induced damage by trapping vacancies generated during displacement cascades, thereby reducing vacancy mobility and suppressing cavity nucleation and growth. For example, Gan \textit{et al.}~\cite{gan2004effect} reported that Hf additions to 316 stainless steel reduced Frank loop size and suppressed void formation under Ni-ion and proton irradiation, while Kato \textit{et al.}~\cite{kato1991effects} observed reduced void swelling in electron-irradiated 316L stainless steel with larger solutes such as Hf and Zr. These observations are consistent with the broader view that sufficiently oversized solutes can act as effective vacancy traps, and motivate defect engineering strategies that exploit controlled solute additions to tailor point defect transport.

Recent high-throughput and machine learning investigations have demonstrated that atomic mobility, rather than thermodynamic nobility alone, plays a dominant role in molten salt corrosion. In particular, Goh \textit{et al.} \cite{goh2024nobility} showed that the bulk tracer diffusion coefficient and chemical diffusion coefficient of Ni are the most impactful predictors of corrosion magnitude in CrFeMnNi alloys. Their Shapley additive explanation (SHAP) analysis showed that higher Ni bulk diffusion strongly correlates with increased total corrosion and ligament forming dealloying morphologies, whereas slower Ni diffusion is associated with reduced corrosion and faceted surface attack. Importantly, the authors noted that the bulk tracer diffusion coefficient can serve as a first order proxy for the surface diffusion coefficient, since both processes arise from similar vacancy-mediated atomic mechanisms \cite{goh2024nobility}. Although surface diffusion is typically faster due to reduced bonding constraints at free surfaces, relative mobility trends among alloys are expected to remain correlated. These findings indicate that intrinsic alloy mobility, as captured by bulk diffusion coefficients, is a key descriptor governing corrosion behavior. This strengthens our analysis of defect mobility showing that vacancy trapping by Y and Y$_2$O$_3$ reduces effective atomic transport and may contribute to mitigating corrosion kinetics.

Additional insights are provided by Wang \textit{et al.} \cite{wang2022integrated}, who highlighted that the dissolution rate during molten salt corrosion depends on elemental diffusion from the bulk to the alloy surface. Their results demonstrate that elemental diffusivity within the alloy matrix influences the supply of reactive species to the interface, thereby affecting corrosion progression. Note that while corrosion can sometimes be described as a near-surface phenomenon \cite{liu2021formation,liu2023temperature,arkoub2025surface,arkoub2025percolating,arkoub2026stress}, it remains fundamentally coupled to atomic transport within the bulk, since both bulk transport and near-surface atomic rearrangements are associated with vacancy-mediated processes involved in dealloying and ligament formation. Together, these studies indicate that controlling defect-mediated transport is important to mitigate corrosion kinetics. Therefore, bulk diffusion coefficients serve not simply as empirical correlations, but as physically meaningful indicators of intrinsic defect mobility trends that influence corrosion behavior, as indicated by Goh \textit{et al.} \cite{goh2024nobility}.

The conceptual model reported by Chan \textit{et al.} \cite{chan2025modification} separates three coupled contributions during molten fluoride dealloying: defect-mediated transport of the dissolving element through the alloy to the exposed surface, charge-transfer-controlled dissolution at the metal--salt interface, and transport of dissolved species away from the interface through the salt. These processes operate in series, and the slowest contribution can control the sustained corrosion rate under a given condition. Their study concerns potentiostatically controlled Cr dissolution from Ni--Cr at 600~\textdegree C, whereas the present work examines static exposure of Ni--Mo alloys without intentional Cr addition at 750~\textdegree C. We therefore use this framework to organize the relevant transport and reaction steps, but we do not apply the Cr dissolution mechanism or rate-controlling step from that study to the present system.

For the present alloys, post-exposure ICP-MS shows that Mo is present in the salt, but it does not identify the dissolved Mo species or the interfacial reaction sequence. Salt impurities may provide the chemical driving force for reactions that remove Mo-containing species from the exposed region. Continued attack then requires redistribution of alloy species from the bulk and near-surface matrix toward that region through lattice, grain boundary, or other short-circuit pathways. Substitutional lattice transport in the fcc matrix is vacancy mediated. Therefore, vacancy trapping by Y and Y$_2$O$_3$ may reduce the alloy-side transport contribution by slowing replenishment of species removed near the surface. The present calculations quantify vacancy mobility and trapping in the alloy, but they do not calculate Mo chemical diffusivity, interfacial charge-transfer kinetics, salt-side mass transport, or impurity-specific reactions. Accordingly, the parallel trends in calculated vacancy mobility and measured corrosion are consistent with an alloy-side transport contribution, but they do not define the complete corrosion pathway or identify the rate-controlling step.

Thermodynamic screening at 750~\textdegree C using HSC Chemistry 6.0 \cite{metso2019hsc} provides context for possible interfacial reactions (Table~\ref{tab:hsc_reactions}). Under the selected standard states, reactions of gaseous hydrogen fluoride [HF(g)] with elemental Y, Ni, and Y$_2$O$_3$ are favorable, whereas the listed reactions with metallic Mo, NiO, and MoO$_2$ are unfavorable. These results indicate that Y$_2$O$_3$ should not be assumed to remain chemically inert, but they do not quantify its dissolution in FLiNaK or identify the pathway responsible for the measured Mo concentration. Determining the detailed Mo dissolution mechanism will require measurements of relevant salt impurities, identification of dissolved Mo species, and quantitative post-corrosion composition profiles. This screening does not identify the complete corrosion pathway or its rate-controlling step.

\begin{table}[!htbp]
\centering
\caption{Standard Gibbs energy changes for selected elemental and oxide reactions with gaseous HF at 750~\textdegree C calculated using HSC Chemistry 6.0 \cite{metso2019hsc}. Values are reported for the reactions as written and rounded to the nearest kJ.}
\label{tab:hsc_reactions}
{\begin{tabular*}{0.92\linewidth}{@{\extracolsep{\fill}}lc@{}}
\toprule
Reaction & $\Delta G^\circ$ (kJ) \\
\midrule
$\frac{1}{3}$Y + HF(g) $\rightarrow$ $\frac{1}{3}$YF$_3$ + $\frac{1}{2}$H$_2$(g) & $-274$ \\
$\frac{1}{2}$Ni + HF(g) $\rightarrow$ $\frac{1}{2}$NiF$_2$ + $\frac{1}{2}$H$_2$(g) & $-32$ \\
$\frac{1}{4}$Mo + HF(g) $\rightarrow$ $\frac{1}{4}$MoF$_4$ + $\frac{1}{2}$H$_2$(g) & $+6$ \\
\midrule
Y$_2$O$_3$ + 6HF(g) $\rightarrow$ 2YF$_3$ + 3H$_2$O(g) & $-240$ \\
NiO + 2HF(g) $\rightarrow$ NiF$_2$ + H$_2$O(g) & $+16$ \\
MoO$_2$ + 4HF(g) $\rightarrow$ MoF$_4$ + 2H$_2$O(g) & $+290$ \\
\bottomrule
\end{tabular*}
}
\par\smallskip
\begin{minipage}{0.92\linewidth}
\footnotesize The notation (g) denotes a gas-phase species. Unmarked species follow the condensed-phase convention in HSC Chemistry 6.0. These values provide standard-state thermodynamic screening. They do not include species activities, non-ideal interactions, or reaction kinetics in molten FLiNaK.
\end{minipage}
\end{table}

The screened reactions do not include soluble Mo oxyfluoride complexes or species activities in FLiNaK. Liu \textit{et al.} reported that sufficiently high O$^{2-}$ concentrations promoted selective Mo leaching from Hastelloy N through the formation of soluble Mo oxyfluoride complexes \cite{liu2023oxide}. Ouyang \textit{et al.} further reported that long-term corrosion of Hastelloy N and Hastelloy B3 in moisture-containing FLiNaK was controlled by outward diffusion of Mo and Cr from the matrix after near-surface depletion developed \cite{ouyang2014longterm}. Although the alloy compositions and salt conditions differ from those used here, these studies suggest a possible coupled process in which transport within the alloy supplies Mo to an impurity-dependent interfacial reaction. Because S and O were not measured and the dissolved Mo species were not identified in the present tests, we cannot determine whether this pathway occurred here.

Note that corrosion kinetics may also involve accelerated transport along grain boundaries. A complete mechanistic description will require future simulations that explicitly address grain boundary diffusion in the presence of vacancy trapping and its coupling with evolving interfacial morphology. Nevertheless, the strong correlation between bulk diffusion and corrosion magnitude reported in recent studies \cite{goh2024nobility,wang2022integrated} supports defect mobility as an important contribution to molten salt corrosion kinetics. The parallel trends in calculated vacancy mobility and measured corrosion penetration in the present work provide further support for this contribution. However, the present work does not resolve the relative contributions of lattice, grain boundary, and other short-circuit transport, or determine their importance relative to interfacial reactions and salt-side transport.

Several limitations of the present modeling framework should be acknowledged. First, the SCMF analysis is parameterized using DFT energetics derived primarily from dilute Ni--Y configurations, without explicitly incorporating additional alloying elements such as Cr, Fe, and Mo that are commonly present in commercial Ni-based alloys. In concentrated or multicomponent systems, the vacancy migration barriers and binding energies can be strongly influenced by the local chemical environment. Capturing such effects would require cluster expansion or related local cluster expansion approaches to construct a reliable energy model that accounts for short-range chemical order and configurational variations. Such a treatment would enable a more rigorous description of vacancy energetics and migration barriers in chemically complex matrices to improve quantitative predictive capability beyond the dilute approximation applied in this study. In addition, the experimental alloys are not ideal single-phase solid solutions. Although the matrix is predominantly fcc Ni--Mo, processing-related oxide secondary phases are present \cite{zhang2026ods}. Therefore, the model is used to isolate the possible transport contribution of Y solutes and Y$_2$O$_3$ dispersions, rather than to describe the full local corrosion chemistry of the experimental alloys.

Second, the present calculations assume a homogeneous solid solution of Y and do not account for possible Y clustering or grain boundary segregation. In practice, Y exhibits a strong thermodynamic tendency to segregate to defects such as grain boundaries or to form clusters or oxide particles (e.g., Y$_2$O$_3$). These effects may locally modify the vacancy population and transport pathways. If Y atoms cluster or segregate, the effective solute concentration in the matrix would decrease, potentially modifying the vacancy-trapping effect predicted here. The present predictions may therefore be interpreted as an upper bound for vacancy mobility suppression under ideal dilute solid solution conditions. A comprehensive treatment would require coupling atomistic energetics with segregation thermodynamics and microstructure evolution modeling as the yttrium distribution might not be homogeneous.

Additionally, the possible influence of Y and Y$_2$O$_3$ additions on local electrochemical behavior was not explicitly investigated in this study. Although yttrium is electrochemically active, the concentration employed here (0.05~wt.\%, corresponding to $\sim$0.035~at.\%) is dilute, and Y$_2$O$_3$ is present as an oxide dispersion in the as-fabricated alloy, as reported by Zhang \textit{et al.} \cite{zhang2026ods}. The present data do not separate the contribution of Y-associated interfacial reactions from the alloy-side transport contribution. The high melting point of Y$_2$O$_3$ \cite{schneider1974melting} supports its thermal stability at the test temperature, but does not by itself prove complete chemical stability in FLiNaK. The thermodynamic screening in Table~\ref{tab:hsc_reactions} also indicates that reaction of Y$_2$O$_3$ with HF(g) is favorable under the selected standard states, but it does not quantify dissolution in FLiNaK. The ICP-MS results in Table~\ref{tab:post_test_icpms} show that Y is detectable in the salt from the NMYO test. This observation indicates that the added oxide dispersion should not be assumed to remain chemically unchanged throughout exposure. However, ICP-MS cannot distinguish Y released from matrix solute, oxide particles, or reaction products. The model does not currently consider dissolution of Y$_2$O$_3$ particles at the salt--alloy interface. It describes vacancy trapping by particles within the bulk alloy before they are reached by the corrosion front. More complete post-corrosion mapping and phase analysis, including quantitative electron probe microanalysis (EPMA) of Y and Mo redistribution, would be needed to resolve the local interfacial chemistry in the present alloys. Therefore, the present experiments document the corrosion penetration trend and are consistent with a defect transport contribution, while the post-corrosion chemical state and distribution of Y$_2$O$_3$ particles remain to be characterized further.
The experimental matrix also contains only one Y level and one Y$_2$O$_3$ loading. Therefore, the present experimental results should be viewed as a comparison among three representative alloy states rather than a complete experimental composition series. How corrosion resistance evolves with Y content and Y$_2$O$_3$ volume fraction is of strong interest and should be explored further through controlled composition-dependent corrosion testing. The parallel trends in predicted vacancy mobility and measured corrosion penetration support further investigation of a possible alloy-side transport contribution.

Finally, several coupled phenomena relevant to alloy performance were not explicitly addressed. For example, the influence of Y on grain boundary strength and intergranular cracking, including possible interactions with salt impurities or environmental species, remains beyond the scope of this study. In addition, under irradiation conditions in MSRs, Y and Y-containing oxides may interact strongly with point defects, defect clusters, or dislocations, potentially promoting impurity defect complexes or modifying recombination kinetics \cite{ke2023microstructure,ke2025role}. While the radiation-induced defects or clusters were not considered in this study, such interactions could either modify radiation tolerance or degrade mechanical properties, depending on the defect chemistry and resulting microstructures. These complex effects require integrated experimental investigation and advanced simulation techniques across scales (e.g., molecular dynamics, kinetic Monte Carlo, rate theories, or cluster dynamics \cite{simon2025multiscalea,simon2025multiscaleb,simon2022mechanistic}) to clarify the broad implications of Y additions on defect evolution and long-term alloy stability.

\section{Methods}
\label{sec:methods}
\subsection{Computational framework for diffusion in fcc Ni with dilute Y}
\label{sec:diffusion model}
To investigate Y diffusion in fcc Ni, we combine density functional theory (DFT), climbing image nudged elastic band (CI-NEB) calculations, and self-consistent mean-field (SCMF) modeling. This workflow quantifies vacancy-mediated diffusion while capturing strong solute--vacancy interactions and kinetic correlations that are important for oversized solutes, following our prior SCMF-based diffusion studies with adaptations for the fcc Ni lattice.

Diffusion in dilute fcc alloys is often analyzed using the five-frequency model \cite{allnatt2003atomic,LECLAIRE197870}, which assumes a well-defined lattice vacancy and treats solute--vacancy interactions primarily within the first two neighbor shells. However, this framework can break down for highly oversized solutes. First-principles studies indicate that when a vacancy occupies the first-nearest-neighbor (1nn) position to an oversized solute, strong relaxation can eliminate a distinct lattice-centered vacancy configuration, producing vacancy delocalization and effectively suppressing the conventional solute--vacancy exchange event. For example, Naghavi \textit{et al.}~\cite{NAGHAVI2017467} reported a vanishing $w_2$ barrier for Y in fcc Co and classified Y as a type III solute based on its large Goldschmidt size mismatch \cite{smithells2013metals}. Given the similarly extreme size mismatch of Y in fcc Ni, analogous nonclassical vacancy configurations and longer-range kinetic correlations are expected.

To address these limitations, we adopt the SCMF approach \cite{nastar2005mean}, which provides a systematically improvable treatment of vacancy-mediated transport by incorporating kinetic correlations through cluster-based transport coefficients \cite{schuler2020kineclue}. SCMF has been applied to diffusion and defect transport problems involving oversized solutes and irradiation-driven phenomena \cite{ke2023microstructure,ke2025diffusion,ke2026,messina2014exact,messina2020solute,ke2022cluster}, and is well suited here because it can accommodate nonstandard local defect configurations and solute-induced perturbations beyond the second nearest neighbor.

\subsection{SCMF diffusion model for Y---vacancy interactions in fcc Ni}
\label{sec:scmf}
SCMF calculations are performed using the \textsc{KineCluE} package \cite{schuler2020kineclue}. The package enables the systematic construction and solution of cluster-based diffusion models in crystalline solids. Y diffusion in fcc Ni is treated within the dilute limit, assuming vacancy-mediated transport. By progressively increasing the spatial range of solute--vacancy interactions included in the model, the SCMF transport coefficients converge toward the exact solution for a single solute--vacancy cluster embedded in an infinite crystal. Unlike the conventional five-frequency model, which restricts solute--vacancy interactions to the 1nn and 2nn shells, the present SCMF framework explicitly includes vacancy configurations and jump processes extending beyond 2nn to 6nn. This allows us to quantify the influence of Y on atomic transport over an extended spatial range and to assess the potential contribution of long-range elastic and chemical interactions to diffusion and vacancy mobility.

The diffusion model consists of a network of vacancy jump frequencies modified by the presence of a Y atom. Vacancy-Ni exchange far from the solute is described by the bulk jump frequency, denoted as $\omega_0$, while Y--vacancy exchange is denoted as $\omega_2$. Additional vacancy jumps $\omega_{ij}$ correspond to transitions between sites $i$ and $j$, where $i$ and $j$ label the $i$-th and $j$-th nearest neighbor positions relative to the Y atom. The interaction range is increased until convergence of the tracer diffusion coefficient is achieved, ensuring that all relevant kinetic correlations are captured.

Within transition state theory \cite{eyring1935activated}, the vacancy jump frequency from site $i$ to site $j$ is given by
\begin{equation}
\omega_{ij} = \nu_{ij} \exp\left(-\frac{E^{ij}_{\mathrm{mig}}}{k_B T}\right),
\end{equation}
where $\nu_{ij}$ is the attempt frequency, $E^{ij}_{\mathrm{mig}}$ is the migration barrier obtained from CI-NEB calculations, $k_B$ is the Boltzmann constant, and $T$ is the temperature. $\nu_{ij}$ is assumed to be constant for all jumps as in fcc Ni, which was calculated as 4.63 THz by DFT \cite{KE201930}. A more rigorous estimation of the attempt frequency can be obtained using Vineyard's harmonic transition state theory, which accounts for vibrational modes at both the ground state and the saddle-point configuration. However, previous studies \cite{messina2014exact,KE201930} have shown that adopting a constant, representative attempt frequency provides a reasonable approximation for vacancy-mediated solute diffusion, and that more detailed phonon-based treatments do not significantly change the predicted diffusion trends. Since diffusion behavior is primarily controlled by vacancy migration barriers and solute--vacancy interaction energetics, this approximation is not expected to affect the main conclusions of this work.

For Y in fcc Ni, DFT structural relaxations reveal a diffusion mechanism analogous to that previously reported for oversized fission product solutes in body-centered cubic (bcc) Fe \cite{ke2025diffusion,bocquet2017migration}. When a vacancy occupies the first-nearest-neighbor (1nn) position relative to a Y atom, the relaxed ground-state configuration corresponds to a delocalized vacancy state in which both the Y atom and the vacancy shift toward an intermediate position between the ideal substitutional site and the 1nn lattice site. Consequently, the vacancy no longer remains centered on a well-defined lattice site, but relaxes toward the Y atom, forming an off-lattice equilibrium configuration. This vacancy splitting behavior fundamentally modifies the conventional solute--vacancy exchange mechanism (\(\omega_2\)) assumed in the five-frequency model, as neither the ground state nor the transition state can be described by well-defined lattice site occupations. To account for this effect within the SCMF framework, the Bocquet mechanism \cite{bocquet2017migration} was incorporated into the \textsc{KineCluE} interaction files by representing the relaxed solute--vacancy configuration with an intermediate solute--vacancy position ($(0.25a, 0.25a, 0)$ position, where $a$ is the lattice constant of fcc Ni) along the solute--vacancy axis and using a 1nn binding energy calculated from DFT. This treatment enables a physically consistent description of vacancy-mediated diffusion in the presence of vacancy delocalization. All saddle-point energies, migration barriers, and binding energies calculated from CI-NEB and DFT are then implemented in the \textsc{KineCluE} interaction files \cite{schuler2020kineclue}.

\subsection{First-principles calculations of Y--vacancy interactions in fcc Ni}
\label{sect:DFT}

Following established first-principles methodologies for predicting diffusion properties in metallic systems \cite{KE201930,ke2022first}, all DFT calculations were performed using the Vienna \textit{Ab initio} Simulation Package (VASP) \cite{PhysRevB.47.558,PhysRevB.54.11169}. Spin-polarized calculations were initialized with ferromagnetic ordering appropriate for fcc Ni. The projector-augmented wave (PAW) method was employed \cite{PhysRevB.50.17953,PhysRevB.59.1758}, and a plane-wave cutoff energy of 479~eV was used. Exchange-correlation effects were described using the generalized gradient approximation (GGA) in the Perdew--Burke--Ernzerhof (PBE) parameterization \cite{PhysRevLett.77.3865}. The simulation cells consisted of a $3 \times 3 \times 3$ supercell of fcc Ni with 108 atoms, relaxed to an equilibrium lattice parameter of 3.52~\AA. Brillouin zone sampling was performed using Monkhorst--Pack grids corresponding to a reciprocal space k point spacing of approximately 0.13~\AA$^{-1}$. The electronic self-consistency convergence criterion was set to $10^{-6}$~eV, and atomic positions were relaxed until residual forces were below 0.01~eV/\AA.

Vacancy--solute binding energies between Y and a vacancy up to the 10th nearest neighbor (nn) shell were calculated following the formalism used in Ref.~\cite{ke2026}:
\begin{multline}
E^{\mathrm{Y}}_{\mathrm{vb},\,i\text{-nn}} =
E\big((N-2)\mathrm{Ni} + 1\mathrm{Y} + 1\mathrm{Va}_{i\text{-nn}}\big)_{\Omega}
- E\big((N-1)\mathrm{Ni} + 1\mathrm{Va}\big)_{\Omega'} \\
- E\big((N-1)\mathrm{Ni} + 1\mathrm{Y}\big)_{\Omega}
+ E\big(N\mathrm{Ni}\big)_{\Omega'} ,
\end{multline}
where $E$ denotes the DFT energy, $N$ is the total number of lattice sites in the supercell, Y represents the substitutional yttrium atom, and Va denotes a vacancy located at the $i$-th nearest neighbor position relative to the Y atom. Negative binding energies indicate attractive solute--vacancy interactions. The supercell volumes and shapes for $E\big((N-1)\mathrm{Ni} + 1\mathrm{Y}\big)$ were fully relaxed and denoted by $\Omega$, while those for pure Ni and Ni containing a single vacancy were relaxed separately and denoted by $\Omega'$.

Migration barriers for vacancy-mediated atomic jumps were determined using the CI-NEB method \cite{10.1063/1.1323224,10.1063/1.1329672}. Three intermediate images were used to describe each diffusion pathway, with the highest-energy image driven to the saddle point. All CI-NEB calculations were performed using the same convergence criteria and magnetic configuration as the static DFT calculations. Additional details regarding the DFT and CI-NEB setup can be found in Refs.~\cite{KE201930,ke2022first}.

\subsection{Transport coefficients}
\label{sect:transport_coeff}

Within the self-consistent mean-field (SCMF) formalism, atomic transport is described through Onsager transport coefficients that account for correlated vacancy-mediated motion. In the present work, these coefficients were computed by explicitly including contributions from both isolated vacancies and vacancy--solute pairs. The SCMF approach formulates the Onsager matrix as a weighted sum of cluster transport contributions, where the weights are determined by the equilibrium populations of monomeric defects and solute--vacancy pairs. As a result, the transport coefficients depend on the alloy composition, the partition functions of defect clusters, and the associated kinetic correlation terms, all of which are evaluated within the same SCMF formalism.

The required Onsager coefficients were obtained using the \textsc{KineCluE} code, which provides a direct numerical solution of the SCMF equations for a given defect interaction network. From the calculated Onsager matrix, the vacancy diffusivity \(D_{\mathrm{V}}\) was extracted as \(D_{\mathrm{V}} = L_{\mathrm{VV}} / c^{\mathrm{fix}}_{\mathrm{V}}\). \(L_{\mathrm{VV}}\) is the diagonal Onsager coefficient associated with vacancy transport, and \(c^{\mathrm{fix}}_{\mathrm{V}}\) is the effective vacancy concentration used in the SCMF formalism \cite{messina2020solute}.

The equilibrium vacancy concentration was determined from the vacancy formation free energy, expressed in terms of the formation enthalpy \(E_{\mathrm{vf}}\) and entropy \(S_{\mathrm{vf}}\), as \(c_{\mathrm{V}} = \exp\!\left(-E_{\mathrm{vf}} / k_{\mathrm{B}}T\right)\exp\!\left(S_{\mathrm{vf}} / k_{\mathrm{B}}\right)\). An experimentally measured value of \(S_{\mathrm{vf}} = 5.71\,k_{\mathrm{B}}\) was adopted \cite{KE201930}. Further details regarding the formulation of the Onsager coefficients, the definition of \(c^{\mathrm{fix}}_{\mathrm{V}}\), and the numerical implementation of the SCMF equations are provided in Refs.~\cite{schuler2020kineclue}.

\subsection{Impurity trapping model}
\label{sect:trap_model}
To account for vacancy trapping by impurities that are not solutes, such as yttrium oxide ($\text{Y}_2\text{O}_3$), an impurity trapping model is adopted in this work. This model was originally developed by Damask and Dienes \cite{DAMASK1964797,PhysRev.120.99} and Flynn \cite{flynn1964} to describe vacancy annealing in the presence of immobile impurities that form defect complexes. It was later generalized by Mansur \cite{MANSUR1981375} to evaluate effective diffusion coefficients of point defects in impure materials by incorporating multiple defect processes, including trapping, thermal dissociation, recombination, and defect production under irradiation. In the present study, only vacancy trapping and thermal release are considered, as the focus is on isothermal corrosion kinetics in the absence of irradiation or other sources of excess point defects.

Following Mansur’s formulation, the effective vacancy diffusivity in the presence of traps can be expressed as
\begin{equation}
\label{eq:DV_trap}
D^{\mathrm{eff}}_{\mathrm{V}} = \frac{D_{\mathrm{V}}}{1 + \kappa_{\mathrm{V}} \tau_{\mathrm{V}} c_{\mathrm{trap}}},
\end{equation}
$D_{\mathrm{V}}$ is the vacancy diffusivity calculated by SCMF as described above, $\kappa_{\mathrm{V}}$ is the vacancy capture coefficient, $\tau_{\mathrm{V}}$ is the mean residence time of a vacancy at a trap before thermal release, and $c_{\mathrm{trap}}$ is the concentration of trapping sites, or $f_{\text{trap}}/V_a$ where $f_{\text{trap}}$ is the trap volume fraction and $V_a$ is atomic volume. The capture coefficient is given by $\kappa_{\mathrm{V}} = 4\pi r'_{\mathrm{V}} D_{\mathrm{V}}$, where $r'_{\mathrm{V}}$ is the effective capture radius of the trap. The mean trapping time is estimated as $\tau_{\mathrm{V}} = b^{2}/D_{\mathrm{V}} \exp(E_{\mathrm{B}}/k_{\mathrm{B}}T)$, where $b$ is an atomic length scale distance and $E_{\mathrm{B}}$ is the binding energy between a vacancy and the trap. The capture radius and the characteristic length $b$ are both taken to be equal to the lattice constant (3.52~\AA) as a reasonable approximation.

\subsection{Alloy fabrication and compositions}

Ni--Mo alloys with controlled yttrium additions were fabricated as model materials for diffusion and corrosion analysis. Three compositions were examined: Ni--15Mo (NM), Ni--15Mo--0.05Y (NMY), and Ni--15Mo--0.05Y--1Y$_2$O$_3$ (NMYO), with all compositions given in weight percent (wt.\%). Elemental powders of Ni ($>$99.9\%), Mo ($>$99.98\%), Y ($>$99.9\%), and Y$_2$O$_3$ ($>$99.99\%) were used as starting feedstocks, with Y$_2$O$_3$ having a nominal particle size of 30--45~nm. 

Powder mixtures corresponding to the target compositions were prepared by high-energy mechanical alloying under an argon atmosphere using a planetary ball mill. Milling was performed for 30~h at 325~rpm with a 10:1 weight ratio of powder to milling media, with intermittent pauses and alternating rotation directions to limit temperature rise. This process ensured homogeneous mixing of the constituent elements and uniform dispersion of Y and Y$_2$O$_3$ within the Ni--Mo alloy matrix.

Bulk specimens were fabricated by electric field assisted sintering (EFAS) in a direct-current sintering furnace (DCS-5, Thermal Technology, LLC, USA). Approximately 20~g of mechanically alloyed powder was loaded into a 20~mm diameter graphite die inside an Ar-purged glovebox. To prevent carbon diffusion from the graphite tooling and to facilitate removal after sintering, 0.127~mm thick Ta foils were placed between the powder compact and the die and punch surfaces. The die assembly was wrapped with graphite felt to reduce heat loss and improve temperature uniformity during consolidation. Sintering was carried out in a direct-current sintering furnace under uniaxial pressure with a dwell time of 15~min. For NM and NMY alloys, consolidation was performed at 1100~\textdegree C under 50~MPa. For alloys containing Y$_2$O$_3$, higher temperatures (1175--1200~\textdegree C) and pressures (up to 60~MPa) were applied to promote pore closure. A heating rate of 100~\textdegree C/min was used up to 1000~\textdegree C, followed by 50~\textdegree C/min to the peak temperature. After the temperature hold, samples were cooled in the furnace to room temperature. Cooling occurred naturally inside the EFAS chamber, corresponding to an uncontrolled cooling rate of approximately 200~\textdegree C/min. These EFAS parameters were selected based on prior literature and preliminary experimental trials to achieve dense compacts with relative densities exceeding 99\% \cite{yang2017origin,kamil2023investigation}.

Although the experimental materials were designed as Ni--15Mo based model alloys and contain a predominantly fcc Ni--Mo matrix, Mo is not completely retained in solid solution. The diffusion modeling in this work focuses on Ni--Y and Ni--Y$_2$O$_3$ interactions using pure fcc Ni as a reference system. At the compositions considered, the alloys remain predominantly an fcc Ni--Mo matrix, while Zhang \textit{et al.} reported oxide secondary phases, including MoO$_2$ associated with residual oxygen from powder processing and EFAS consolidation \cite{zhang2026ods}. Note that while Mo and these oxide particles may influence diffusion kinetics quantitatively, the present modeling framework is designed to isolate the specific mechanisms associated with Y solutes and Y$_2$O$_3$ dispersions, which are expected to cause vacancy trapping and diffusion suppression.

\subsection{Corrosion testing and microstructural characterization}
Static corrosion tests were performed in molten FLiNaK salt (46.5--11.5--42 mol\% LiF--NaF--KF) at 750~\textdegree C for 500~h to evaluate the degradation behavior of Ni--Mo alloys with and without Y or Y$_2$O$_3$ additions. LiF (99.99\% trace metal basis), NaF (99.99\% trace metal basis), and KF (99.97\% trace metal basis) were procured from Sigma-Aldrich and stored in an Ar-filled glovebox (O$_2$ $<$ 0.1~ppm, H$_2$O $<$ 0.1~ppm). The FLiNaK mixture was prepared in-house using a balance with $\pm$0.0001~g precision. The FLiNaK salt was dried before testing and handled under inert atmosphere to minimize moisture and oxygen contamination. To further reduce impurities, the salts were thermally purified by baking prior to testing. After melting, any gray floating material was removed, and the purified salt was cast into a glassy carbon boat, cooled, broken into chunks, sealed in glass vials, and stored in the glovebox until use.

Approximately 10~g of salt was loaded into a glassy carbon crucible, which was placed inside a secondary graphite crucible to minimize contamination and maintain chemical stability at elevated temperature. Specimens were mounted on a boron nitride lid using Ni wire to avoid direct contact with the crucible and to eliminate possible galvanic effects. The furnace was heated from room temperature to 750 \textdegree C, and samples were immersed once the temperature stabilized. For the NM, NMY, and NMYO comparison reported in this manuscript, two specimens of each composition were exposed under the same molten salt condition. To minimize variability associated with salt volume, each crucible contained approximately the same mass of salt ($\sim$10~g). In addition, thermal exposure controls were performed for each material by heating specimens in the same furnace without salt.

At the final stage of testing, specimens were removed from the molten salt and allowed to cool to room temperature. Residual salt was removed by repeated ultrasonic cleaning in deionized water and ethanol until no further mass change was detected. Specimen masses were recorded before and after corrosion testing. The weight change ratio was calculated as $\Delta m/m_0 \times 100\%$, where $m_0$ is the initial mass and $\Delta m=m_f-m_0$. Negative values indicate net mass loss after cleaning. Microstructure characterization after corrosion was conducted using scanning electron microscopy (SEM) to evaluate surface degradation and cross-sectional corrosion attack depth. Corrosion penetration depth was estimated from cross-sectional SEM observations based on surface recession, corrosion porosity, and near-surface morphological features. The maximum penetration depth was estimated as the distance from the visible remaining exposed surface to the deepest continuous region of corrosion porosity, recession, or near-surface attack. Because the original interface before corrosion was not independently marked, these values are reported as approximate ranges. The average penetration depth for each specimen was estimated from repeated line measurements across the 500X overview SEM images. The mean penetration reported in Table~\ref{tab:corrosion_metrics} represents the mean and one standard deviation calculated from the average penetration depths from the two specimens. Individual specimen values and line measurement statistics are provided in Table~\ref{tab:corrosion_line_measurements}. This analysis from SEM images was used to quantify the extent of corrosion damage and directly compare degradation behavior among the tested alloy compositions.
For selected corrosion tests, salts after exposure were sampled and analyzed by inductively coupled plasma mass spectrometry (ICP-MS) to quantify selected metallic species in the salt. Sulfur and oxygen concentrations were not measured directly in the starting or post-exposure salts. Standard Gibbs energy changes for the selected reactions in Table~\ref{tab:hsc_reactions} were calculated at 750~\textdegree C using HSC Chemistry 6.0 \cite{metso2019hsc}. The reactions were evaluated as written under standard-state conditions without species activities, non-ideal salt interactions, or kinetic corrections.

\section*{Acknowledgements}
This work was supported by the Laboratory Directed Research and Development (LDRD) Program at Idaho National Laboratory. The funder supported the research but had no role in the analysis, interpretation, decision to publish, or preparation of the manuscript. This manuscript has been authored by Battelle Energy Alliance, LLC under Contract No.~DE-AC07-05ID14517 with the U.S. Department of Energy. This research made use of the resources of the High Performance Computing Center at Idaho National Laboratory, which is supported by the Office of Nuclear Energy of the U.S. Department of Energy and the Nuclear Science User Facilities under Contract No.~DE-AC07-05ID14517. The United States Government retains, and the publisher, by accepting the article for publication, acknowledges that the United States Government retains a nonexclusive, paid-up, irrevocable, worldwide license to publish or reproduce the published form of this manuscript, or allow others to do so, for United States Government purposes.

\section*{Data availability}
The data that support the findings of this study are available from the corresponding author upon reasonable request.

\section*{Competing interests}
The authors declare no competing financial or non-financial interests.

\section*{Contributions}
J.-H.K. – Conceptualization, formal analysis, investigation, methodology (lead), visualization, data curation, writing – original draft, writing – review and editing. X.Z. – Investigation, methodology, formal analysis, manuscript review. J.Y. – Investigation, methodology, formal analysis, manuscript review. Q.Y. – Investigation, methodology, formal analysis, manuscript review. R.G. – Investigation, methodology, formal analysis, manuscript review. R.S. – Conceptualization, investigation, methodology, funding acquisition, supervision, writing – review and editing.





\nocite{schneider1974melting}
\bibliography{sn-bibliography.bib}

\end{document}